\documentclass[aps,twocolumn,pra]{revtex4}
\usepackage[dvips]{graphicx}
\newcommand\beq{\begin{equation}}
\newcommand\eeq{\end{equation}}
\newcommand\bea{\begin{eqnarray}}
\newcommand\eea{\end{eqnarray}}

\begin{document}

\title{Pathways to non-sequential multiple ionization in strong laser fields}

\author{Krzysztof Sacha\dag\ and Bruno Eckhardt\ddag  
}

\affiliation{\dag\ Instytut Fizyki im. Mariana Smoluchowskiego,
Uniwersytet Jagiello\'nski, ul. Reymonta 4,
PL-30-059 Krak\'ow, Poland \\
\ddag\ Fachbereich Physik, Philipps-Universit\"at Marburg,
D-35032 Marburg, Germany}

\begin{abstract}
The outgoing electrons in non-sequential multiple ionization 
in intense laser fields are strongly correlated. 
The correlations can be explained within a classical model
for interacting electrons in the presence of
the external field. Here we extend the previous analysis
for two and three electrons to cases with up to eight electrons
and identify the saddle configurations that guard the
channels for non-sequential multiple ionization.
For four and fewer electrons the electrons
in the dominant configuration are equivalent, 
for six and more electrons this is no longer the case. 
The case of five electrons is marginal,
with two almost degenerate transition configurations. The total 
number of configurations increases rapidly, from 2 configurations
for three electrons up to 26 configurations for eight electrons.
\end{abstract}
\maketitle


\section{Introduction}
Laser pulses with peak intensities of the order of $10^{14}$~W/cm$^2$ 
and wavelengths in the infrared can ionize several electrons
simultaneously \cite{luk83,huillier83,boyer84}.
Growing experimental evidence 
\cite{walker94,moshammer00,weber00prl,weber00jpb,weber00n,wechenbrock01,
feuerstein01,moshammer02}
supports a rescattering
process \cite{corkum93,kulander93}
as the most likely mechanism for the rapid
transfer of a sufficient amount of energy to the electrons. 
The interaction between the rescattered electron and the others 
takes place close to the nucleus, leading to the formation
of a short lived, highly excited compound state with electrons
near to the nucleus. This compound state will decay by emitting
a single electron (the most likely decay mode) or several electrons.
The highly excited electrons move so quickly that the
changes in the field can be neglected: 
the electrons see an essentially static electric field.
Experimental studies of the final momentum distribution for double ionization
show that the electron momenta parallel to the field are strongly correlated
\cite{weber00n,wechenbrock01,feuerstein01,moshammer02}. 
Building on the picture of an intermediate compound state 
we have proposed a
classical origin for the electron correlations
\cite{eckhardt01,eckhardt01pra1,eckhardt01silap,eckhardt01pra2,eckhardt01epl}. 
If the energy
of the compound state is above the zero field threshold energy
no constraints on the outgoing dynamics are imposed. If the energy
is below double ionization can occur only if the field is still
present. Then the electrons will aim for the Stark saddle, but
the electron repulsion will prevent them from crossing the saddle
in the same place at the same time. Furthermore, if one electron is ahead it
will push the others back, thus preventing
multiple ionization. Therefore, the electrons have
to escape in a configuration that balances the mutual repulsion.
For two electrons this means that the electrons move 
side by side with respect to the field axis. 
Within this symmetric subspace it is easy to identify a saddle 
which the electrons have to cross on their way to double ionization: 
it is a stationary configuration of two electrons side by side
close to the Stark saddle for a single electron
(assuming, as mentioned before, that the field is stationary).
The number of saddle configurations increases with the number
of electrons: there are two configurations for three electrons, 
with different energies and different threshold
exponents \cite{eckhardt01pra2}.

We here extend our earlier study of such saddle configurations for
two and three electrons to four or more. The aim is to identify the
possible saddles, their critical energies and their threshold 
exponents. The number of configurations increases rapidly with the
number of electrons. Beginning with five electrons also several
nearby almost degenerate saddles appear. Most interestingly,
for four and fewer electrons all electrons are equivalent in the
dominant mode, but this is no longer true for six or more.

We conclude this introduction with notation and the formulation of
the Hamiltonian. Since the electrons escape from the highly excited
compound state so quickly, we may assume that the field is 
constant during the escape. Let 
the electric field point in the $z$-direction and let the
electrons be labeled $i=1,\ldots,N$ with positions 
${\bf r}_i=(x_i,y_i,z_i)$.
Then the Hamiltonian is, in atomic units,
\beq
H_N = \sum_{i=1}^N \frac{{\bf p}_i^2}{2}- \sum_{i=1}^N \frac{Z}{|{\bf r}_i|}
+ \sum_{i<j}^N\frac{1}{|{\bf r}_i-{\bf r}_j|} - F\sum_{i=1}^N z_i\,.
\label{ham}
\eeq
Simple scaling shows that the field strength $F$ can be absorbed, so
that without loss of generality we can take $F=1$
\cite{eckhardt01pra1,eckhardt01pra2}.

The aim now is to determine the stationary points of this Hamiltonian,
where the derivatives with respect to momenta and positions vanish. 
For the most
part this is not possible analytically. The numerical method of choice
is the Newton-Raphson iteration
\cite{numericalrecipes}. The required calculation of the 
matrix of second derivatives can be turned to an advantage
since it allows for an immediate determination of the stability
properties of configuration. The main disadvantage of the method,
but also of all other numerical methods, is that they do not provide
any measure by which to judge whether all solutions have been found.
On the practical side we can make sure that we have hit at least 
all the configurations with
large domains of attraction in the Newton-Raphson method by starting
with a sufficiently large number of randomly selected initial
conditions. We used up to $10^6$ initial conditions. The
fact that fairly similar but definitely different configurations 
could be detected may be interpreted as strong evidence that 
all configurations have indeed been found.

The stability analysis of the stationary point in the 
full phase space allows us
to determine the behavior of the cross-section for non-sequential
escape close to the threshold \cite{eckhardt01epl}. 
Among all Lyapunov exponents of the saddle there is one, 
$\lambda_r$, whose eigenvector components point in the same
direction along the field axis: it corresponds to a simultaneous 
motion of all electrons in the same direction away from the
saddle. Borrowing terminology from chemical reactions, 
we call this subspace the reaction coordinate. Because
of the repulsion between electrons all configurations have
additional unstable eigenspaces, which enter in the
threshold exponent.
If the initial energy of the system equals the saddle energy
only a trajectory living in the symmetric subspace can lead to
a simultaneous escape of all electrons. This reduces the dimensionality of the
problem and the cross-section vanishes. For higher energy some deviations from
the symmetric motion are possible, giving a finite volume of
initial conditions and a non-vanishing cross-section.
The dependence of the cross-section on energy $\sigma(E)$ close to the saddle 
energy $E_N$ can be obtained in the spirit of the Wannier analysis
\cite{wannier53,rau84,rost98,rost01phe}, resulting in
\beq
\sigma(E) \sim (E-E_N)^\mu,
\eeq
with an exponent
\beq
\mu=\frac{1}{\lambda_r}\sum_{i=1}^{n_u}\lambda_i \,.
\eeq
The $\lambda_i$'s are the positive Lyapunov exponents of the saddle except for
the reaction coordinate exponent $\lambda_r$. 
The cross section is large if the exponent is small, i.e. if 
the saddle is crossed quickly (large $\lambda_r$)
or if the differences from the symmetric motion grow slowly
(small $\lambda_i$).
These cross section exponents are an additional characteristic
of the multiple ionization process.

The remainder of the paper is organized as follows: in the next
section we discuss a few obvious, and highly symmetric configurations.
This is followed by our numerical results for up to eight electrons
in section~\ref{sec3}. We conclude with a few remarks in section~\ref{sec4}.

\section{Symmetric configurations}
\label{sec2}

\subsection{All electrons on a ring}

\begin{figure}
\begin{center}
\includegraphics*[width=8.6cm]{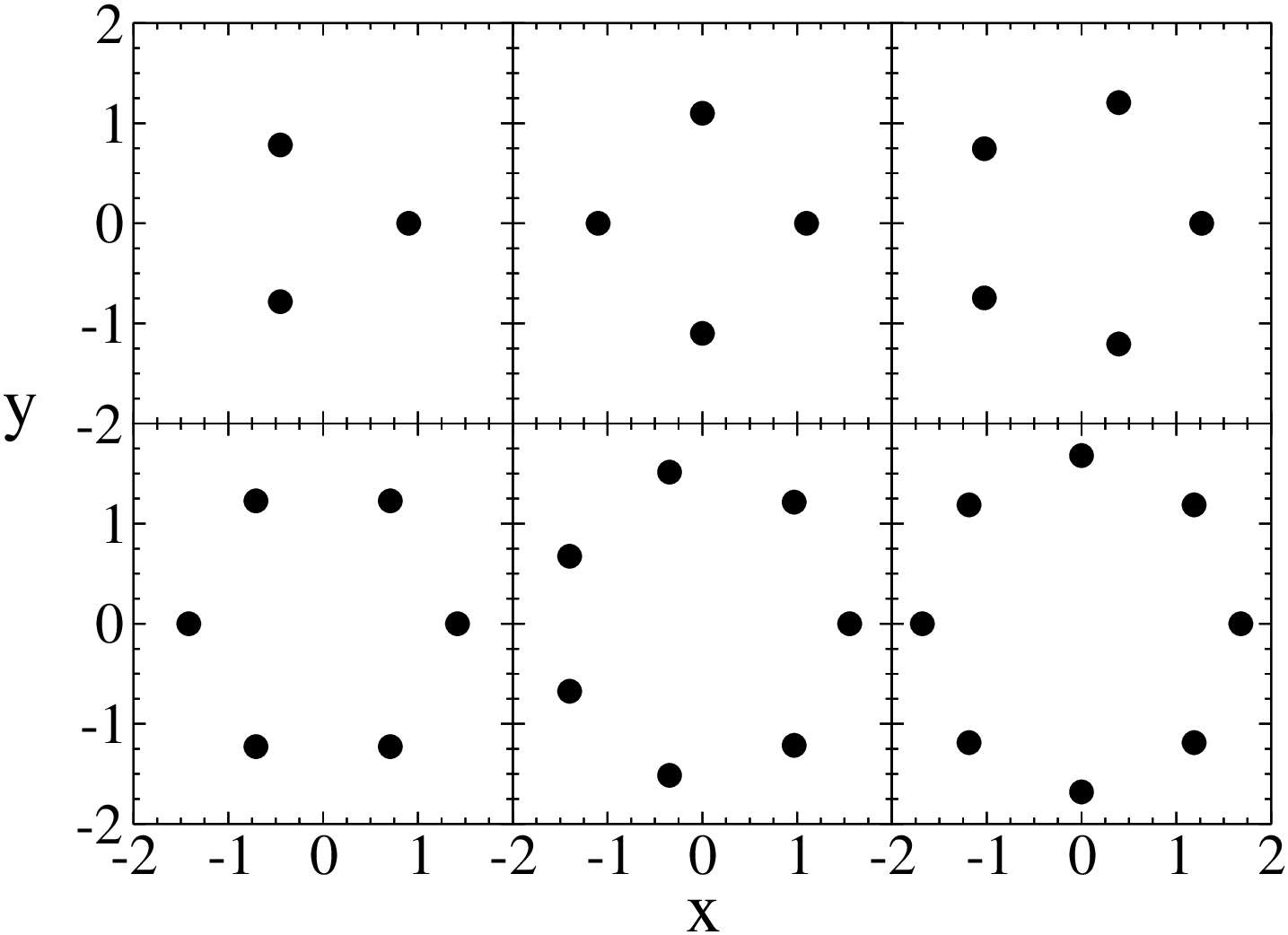}
\end{center}
\caption[]{\label{symmetric_figure}
Saddles with electrons on a ring. Shown are the
cases with $N$ from 3 up to 8, projected
onto the $x$-$y$-plane perpendicular to the field. Note
the increasing spacing between the electrons that indicates
the growing influence of the electron repulsion.
}
\end{figure}

We assume that all electrons are situated in a plane perpendicular to the
field and that they obey a $C_{Nv}$ symmetry. The reflection symmetry 
limits the momenta to be parallel to the symmetry planes and thus 
confines the motion to a dynamically allowed 
subspace in the high-dimensional $N$-body phase space. That is,
if in the full phase space of the $N$-body problem initial
conditions are prepared in this subspace they will never leave it.
In the symmetry subspace we can use
cylindrical coordinates for the electrons,  
$z_i=z$, $\rho_i=\rho$ and $\varphi_i=2\pi i/N$.
Then, for zero total angular momentum along the field axis 
and in the scaled variables where $F=1$, the Hamiltonian (\ref{ham})
can be reduced to 
\beq
H(p_\rho,p_z,\rho,z)=
\frac{p_\rho^2+p_z^2}{2N}-\frac{N^2}{\sqrt{\rho^2+z^2}}
+\frac{W}{2\rho}
-Nz,
\label{hcNv}
\eeq
with 
\beq
W=\sum_{i<j}^{N}\frac{1}{\sin\left[\frac{\pi}{N}(j-i)\right]}
=\sum_{k=1}^{N-1}\frac{N-k}{\sin\left[\frac{\pi k}{N}\right]}\,.
\eeq
The Hamiltonian (\ref{hcNv}) possesses a stationary point at $p_\rho=0$,
$p_z=0$, $\rho=\rho_N$ and $z=z_N$ where
\bea
\rho_N&=&\sqrt{\frac{W}{2N}}
\left[\left(\frac{2N^2}{W}\right)^{2/3}-1\right]^{1/4} \cr
z_N&=&\sqrt{\frac{W}{2N}}
\left[\left(\frac{2N^2}{W}\right)^{2/3}-1\right]^{3/4}\,. 
\eea
and
\bea
E_N&=&-\left[2N^2\left(\frac{2}{NW}\right)^{1/6}-\sqrt{2NW}\right] \cr
&& \left[\left(\frac{2N^2}{W}\right)^{2/3}-1\right]^{-1/4}\,.
\label{ENcNv}
\eea
The first few saddles with electrons on a ring are shown in 
Fig.~\ref{symmetric_figure}.

For $k\ll N$ the repulsion energy $W$ behaves like $W=(N^2/\pi)\sum (1/k)$.
This suggests that $W$ increases like $N^2 \ln N$, and indeed
numerical evidence supports
\beq
W \approx (0.3 N^2 + 0.3 N - 3.1) \ln N\,.
\eeq
Since $W$ increases faster than quadratically, the configurations exist
for finite $N$ only. Specifically, we find that the energy of the 
configuration first decreases but then increases again, and that 
there is no configuration for $N=473$ or more particles
\footnote{This discussion corrects the one given in
\cite{eckhardt01},
where a much smaller number of possible configurations
was obtained using $W=N(N-1)/\sin(\pi/N)$, 
with the large $N$ behavior
$W\sim N^3$. This
overestimate of the repulsion between the
electrons is responsible for the smaller number of possible
configurations.}.
The properties of the configurations, including 
stability and cross section exponents, are 
listed in table~\ref{symmetric}.

\begin{table}
\caption[]{\label{symmetric} Saddles with electrons on a ring in 
a plane perpendicular to the field axis. The columns are the number $N$ 
of electrons, the energy $E_N$ at the saddle, the number $n_u$ of unstable
directions (excluding the one along which all electrons escape, i.e. the 
reaction coordinate), the Lyapunov exponent $\lambda_r$ along the reaction 
coordinate, the critical exponent $\mu$ near threshold, and the radius
$\rho_N$ and the distance $z_N$ of the saddles from the nucleus.}
\begin{center}
\begin{tabular}{rrrrrrr}
\hline
$N$ & $E_N$ & $n_u$ & $\lambda_r$ & $\mu$ & $\rho_N$ & $z_N$\\
\hline
 2 & $ -4.5590$ &  1 &   1.2139 &   1.2918 & 0.6580 & 1.1398 \\ 
 3 & $ -7.6673$ &  2 &   1.1054 &   2.6226 & 0.9036 & 1.2779 \\ 
 4 & $-11.1059$ &  3 &   1.0340 &   4.0971 & 1.0994 & 1.3882 \\ 
 5 & $-14.8004$ &  4 &   0.9817 &   5.7342 & 1.2677 & 1.4800 \\ 
 6 & $-18.7044$ &  8 &   0.9409 &   8.6800 & 1.4175 & 1.5587 \\
 7 & $-22.7859$ & 10 &   0.9077 &  12.0849 & 1.5538 & 1.6276 \\ 
 8 & $-27.0208$ & 12 &   0.8799 &  15.9050 & 1.6794 & 1.6888 \\ 
\hline
\end{tabular}
\end{center}
\end{table}


\subsection{One electron in the center}

\begin{figure}
\begin{center}
\includegraphics*[width=8.6cm]{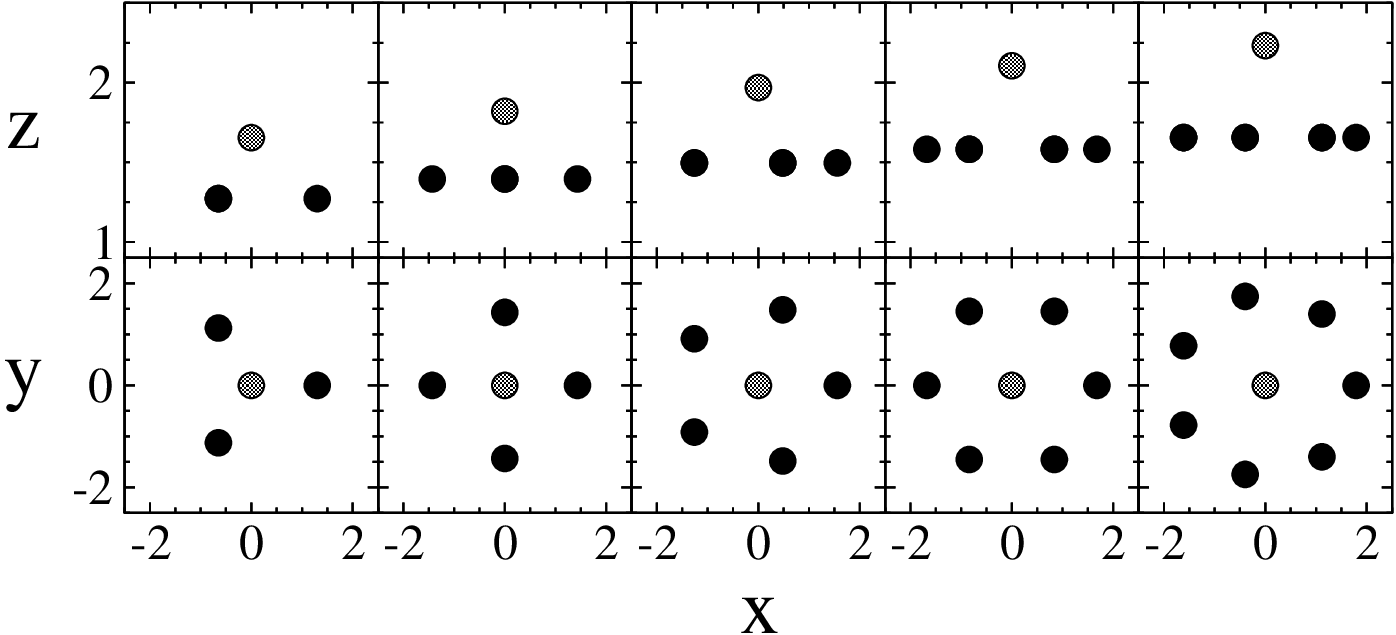}
\caption[]{\label{symmetric_w_center_figure}
Saddles with one electron on the field axis and the others 
on a ring. The electron in the center is shown shaded.
Shown are cases $N=4$, $5$, $6$, $7$ and $8$, 
projected onto the $x$-$z$-plane with the field axis along $z$ (top row) 
and onto the $x$-$y$-plane perpendicular to the field (bottom row). 
}
\end{center}
\end{figure}

The simplest extension from the symmetric ring is to place
one electron on the field axis and the others on a ring as shown in 
Fig.~\ref{symmetric_w_center_figure}.
Such configurations have three characteristic lengths, the 
radius of the ring $\rho_{N-1}$, the position of the ring $z_{N-1}$ and the 
position of the electron on the axis $z_c$. No analytical solutions
have been found. The numerically determined characteristics are
collected in table~\ref{symmetric_w_center}.

\begin{table}
\caption[]{\label{symmetric_w_center}Saddles with one electron on
the field axis and the other $N-1$ electrons on a ring.
The columns are the number $N$ of electrons,
the energy $E_N$ at the saddle, the number $n_u$ of unstable
directions (excluding the one along the reaction coordinate),
the Lyapunov exponent $\lambda_r$ along the reaction coordinate,
the critical exponent $\mu$ near threshold, the position $z_c$ 
of the electron on the axis and the radius
$\rho_{N-1}$ and the distance $z_{N-1}$ from the nucleus for the ring of
$N-1$ electrons.}
\begin{center}
\begin{tabular}{rrrrrrrr}
\hline
$N$ & $E_N$ & $n_u$ & $\lambda_r$ & $\mu$ & $z_c$ &$\rho_{N-1}$ & $z_{N-1}$\\
\hline
 4 & $-10.9398$ &  3 &   1.0271 &   4.2423 & 1.6543 & 1.2996 & 1.2718 \\ 
 5 & $-14.8001$ &  4 &   0.9750 &   5.6633 & 1.8199 & 1.4324 & 1.3950 \\ 
 6 & $-18.8975$ &  5 &   0.9345 &   7.2035 & 1.9690 & 1.5583 & 1.4960 \\ 
 7 & $-23.1867$ &  6 &   0.9015 &   8.8984 & 2.1054 & 1.6775 & 1.5813 \\ 
 8 & $-27.6372$ &  9 &   0.8739 &  10.9698 & 2.2317 & 1.7905 & 1.6553 \\ 
\hline
\end{tabular}
\end{center}
\end{table}


\subsection{All electrons on a line}
\begin{figure}
\begin{center}
\includegraphics*[width=8.6cm]{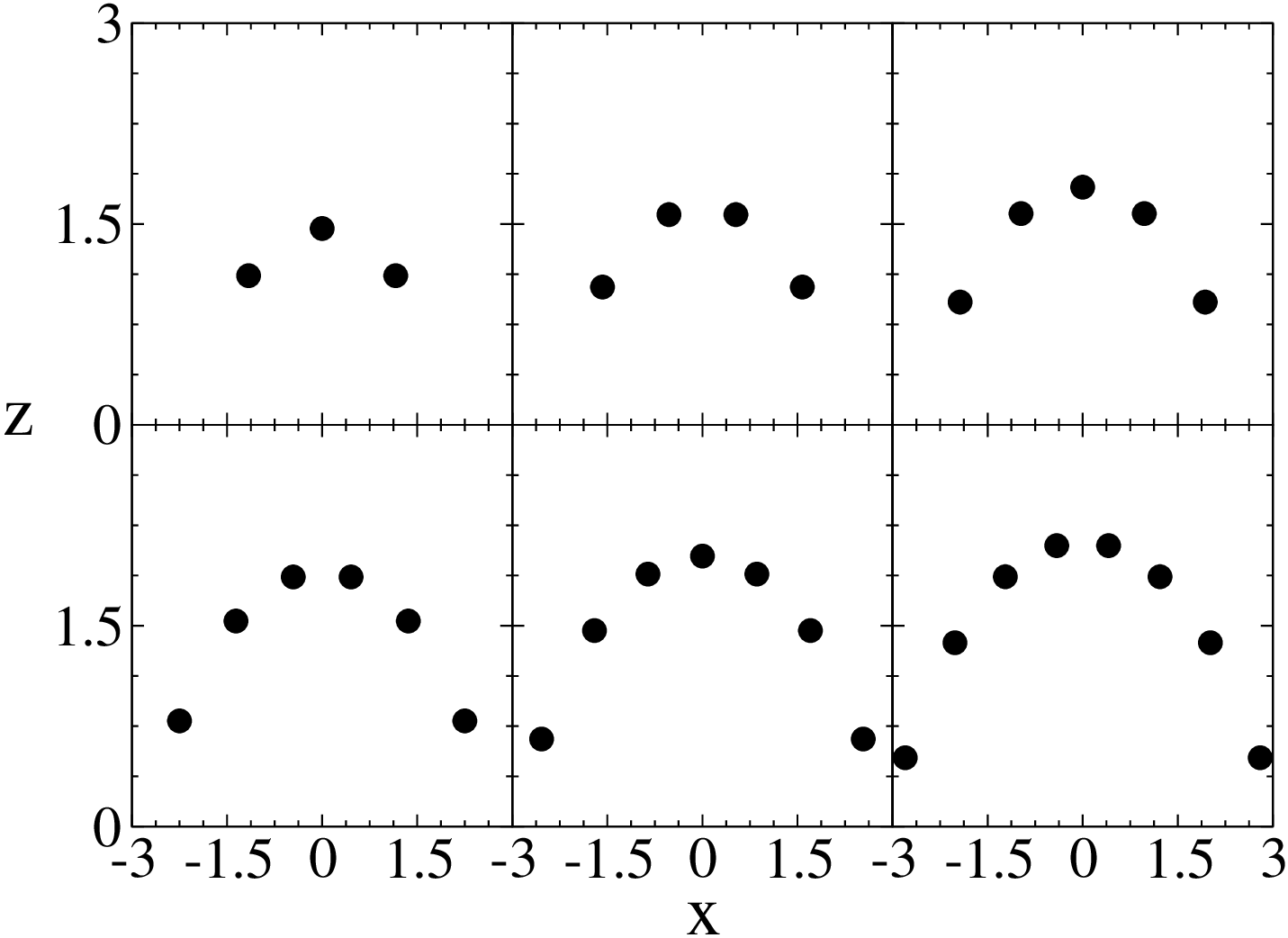}
\end{center}
\caption[]{\label{lineup_figure}
Saddles with all electrons in a plane. 
Shown are cases $N=3$, $4$, $5$, $6$, $7$ and $8$, 
projected onto the $x$-$z$-plane, with the field pointing along $z$ axis. 
}
\end{figure}

A third class of configurations has all electrons in a
plane through the field axis. The electrons are placed
like beads on a string. With the confinement of electrons to
an almost linear arrangement the saddle energies are higher than
in the previous cases and the cross section exponents are also
much larger. The properties of the saddles are listed in table~\ref{lineup},
and their positions are shown in Fig.~\ref{lineup_figure}.

\begin{table}
\caption[]{\label{lineup}Saddles with all electrons in 
a plane through the field axis.
The columns are the number $N$ of electrons,
the energy $E_N$ at the saddle, the number $n_u$ of unstable
directions (excluding the one along the reaction coordinate),
the Lyapunov exponent $\lambda_r$ along the reaction coordinate,
the critical exponent $\mu$ near threshold, and the 
minimal and maximal positions along the field axis,
$z_{min}$ and $z_{max}$, respectively.}
\begin{center}
\begin{tabular}{rrrrrrr}
\hline
$N$ & $E_N$ & $n_u$ & $\lambda_r$ &$\mu$ & $z_{min}$ & $z_{max}$ \\
\hline
 3 & $ -7.3902$ &  3 &   1.0981 &   3.7040 & 1.1143 & 1.4666  \\ 
 4 & $-10.3975$ &  5 &   1.0177 &   6.8831 & 1.0295 & 1.5699  \\ 
 5 & $-13.5302$ &  7 &   0.9568 &  10.7953 & 0.9170 & 1.7755  \\ 
 6 & $-16.7566$ &  9 &   0.9082 &  15.4282 & 0.7898 & 1.8639  \\ 
 7 & $-20.0549$ & 11 &   0.8680 &  20.7764 & 0.6543 & 2.0199  \\ 
 8 & $-23.4091$ & 13 &   0.8339 &  26.8366 & 0.5141 & 2.0986  \\ 
\hline
\end{tabular}
\end{center}
\end{table}

\section{All configurations}
\label{sec3}

\begin{figure}
\begin{center}
\includegraphics*[width=8.6cm]{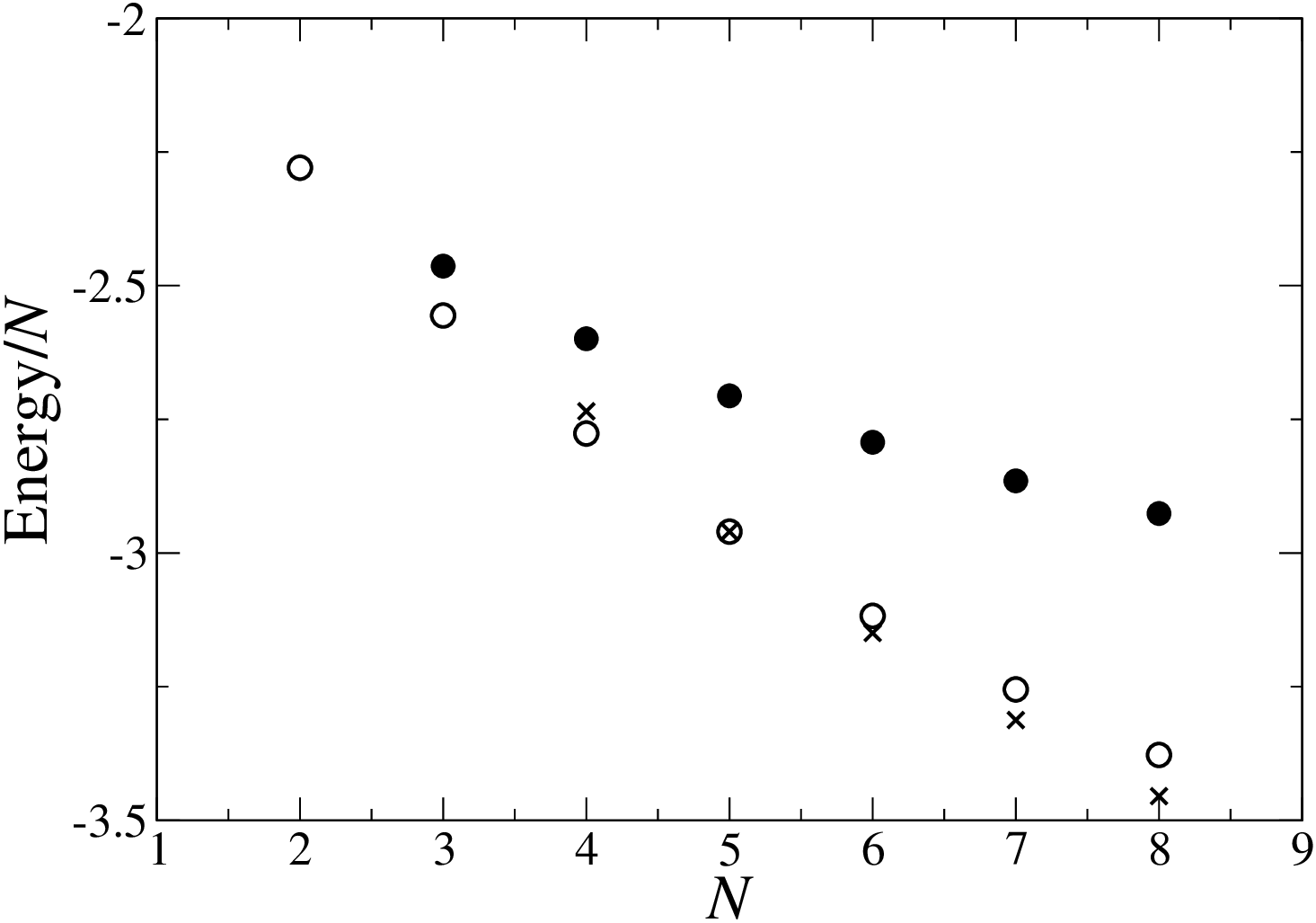}
\end{center}
\caption[]{\label{summary}
Energy of the saddles divided by $N$ in the different symmetry classes
for increasing number of electrons. The open circles are
for electrons on a ring ($C_{Nv}$ symmetry), the crosses for configurations
with one electron on the axis and the others on a ring ($C_{(N-1)v}$ symmetry) 
and full circles are for the configurations with all electrons
on a line ($C_{v}$ symmetry).
}
\end{figure}

A summary of the symmetric configurations discussed in the previous
section is provided in Fig.~\ref{summary},
which shows the saddle energies for different $N$. For the range of 
$N$ included in the figure the saddle energies for all 
configurations decrease, but as the example of the ring configurations
shows, they can be expected to increase for sufficiently large
$N$. There is a crossover in the configuration that gives the minimum,
from the ones with all particles on a ring for $N<5$ to the ones
with a one electron in the center for $N>5$. The case $N=5$ is marginal
(see below). In order to see what other configurations one has to 
consider, we turn to a discussion of configurations 
found with the Newton-Raphson method for 
electron numbers between four and eight. The cases of two and three 
electrons have been analyzed previously
\cite{eckhardt01,eckhardt01pra1,eckhardt01silap,eckhardt01pra2,eckhardt01epl}.

\subsection{Four electrons}
\begin{figure}
\begin{center}
\includegraphics*[width=8.6cm]{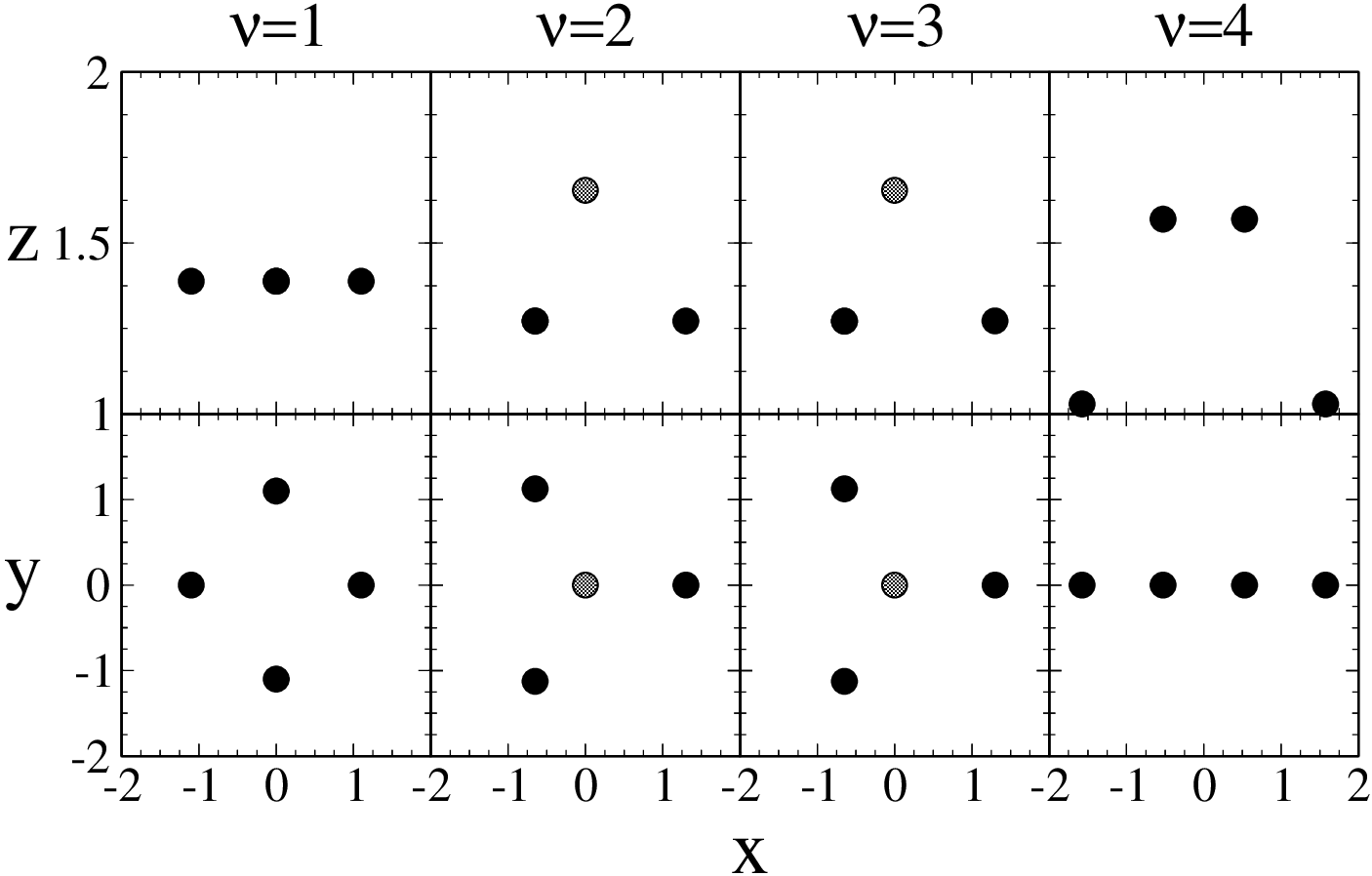}
\end{center}
\caption[]{\label{all_four_fig}
All saddles for four electrons, projected
onto the $x$-$z$-plane with the field axis (top row)
and onto the $x$-$y$-plane perpendicular to the field. 
Some electrons are shown shaded in order to help identify
them in the two sets of frames.
The states have $C_{4v}$ ($\nu=1$), $C_{3v}$ 
($\nu=2$), $C_{v}$ ($\nu=3$) and $C_{2v}$ ($\nu=4$) symmetries.
}
\end{figure}

For four electrons there are 4 configurations, shown in 
Fig.~\ref{all_four_fig}. Their properties are listed in table~\ref{all_four}.
The dominant non-sequential ionization takes place in the vicinity of the 
saddle corresponding to all electrons on a ring in a plane perpendicular 
to the field axis, as in the cases for two and three electrons. 
This saddle possesses both the lowest energy and the smallest critical 
exponent and thus the most favorable dependence of the cross-section on energy. 

The next two configurations are almost degenerate in energy and have similar
critical exponents. They form a triangle with an additional electron 
placed in the
center. The saddle $\nu=2$ possesses a $C_{3v}$ symmetry, but the state
$\nu=3$ has a $C_{v}$ symmetry only. The last $\nu=4$ configuration, with
all electrons in a plane, has significantly higher energy and larger
critical exponent compared to the previous states.

\begin{table}
\caption[]{\label{all_four}
Saddles for four electrons, ordered by increasing saddle energy.
The columns give the number $\nu$ of the state,
the energy $E_\nu$ at the saddle, 
the number $n_u$ of unstable directions 
(excluding the one along the reaction coordinate),
the Lyapunov exponent $\lambda_r$ along the reaction coordinate,
the critical exponent $\mu$ near threshold, and some comment
on the states.}
\begin{center}
\begin{tabular}{rrrrrl}
\hline
$\nu$ & $E_\nu$ & $n_u$ & $\lambda_r$ & $\mu$ & comment\\
\hline
 1 & $-11.1059$ &  3 &   1.0340 &   4.0971 &  all on a ring \\ 
 2 & $-10.9398$ &  3 &   1.0271 &   4.2423 &  a ring plus center \\ 
 3 & $-10.9398$ &  4 &   1.0271 &   4.2708 &  --- \\ 
 4 & $-10.3975$ &  5 &   1.0177 &   6.8831 &  all on a line \\ 
\hline
\end{tabular}
\end{center}
\end{table}

\subsection{Five electrons}
\begin{figure}
\begin{center}
\includegraphics*[width=8.6cm]{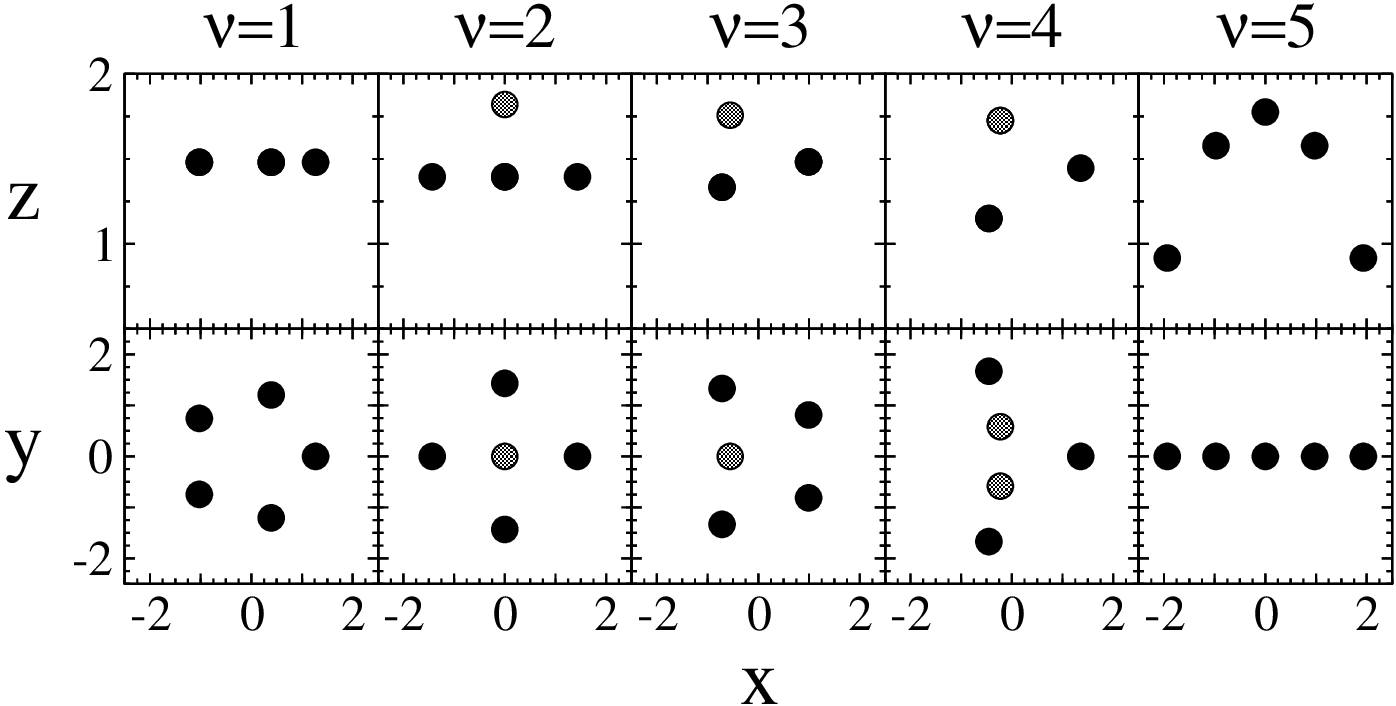}
\end{center}
\caption[]{\label{all_five_fig}
All saddles for five electrons, projected
onto the $x$-$z$-plane with the field axis (top row)
and onto the $x$-$y$-plane perpendicular to the field. 
Some electrons are shaded to help identification 
in the two sets of frames.
The states possess $C_{5v}$ ($\nu=1$), $C_{4v}$ ($\nu=2$), 
$C_{v}$ ($\nu=3,4$) and $C_{2v}$ ($\nu=5$) symmetries.
}
\end{figure}

For five electrons we find 5 configurations shown in 
Fig.~\ref{all_five_fig}, with properties given in table~\ref{all_five}.
The five electron problem constitutes actually a marginal case where the
configuration of all electrons on a ring starts loosing its dominance. 
The energy of this configuration is only slightly lower than the energy of 
the state $\nu=2$ with one electron in the center and the others on a ring. 
The state $\nu=2$ possesses, however, a smaller critical exponent than the 
$\nu=1$ configuration. Therefore, unless there is a wide disparity
in prefactors in the cross section law or residual correlations
in the compound state that prefer one state over the other,
we expect that both will contribute significantly to
the process of non-sequential five electron escape.

The configurations $\nu=3$ and $\nu=4$ have the $C_v$ symmetry and reveal 
higher energy and larger critical exponents than the previous ones. 
The last $\nu=5$ configuration with all electrons on a line is the least 
significant because of its high energy and its high critical exponent.

\begin{table}
\caption[]{\label{all_five} 
Saddles for five electrons, ordered by increasing saddle energy.
The columns give the number $\nu$ of the state,
the energy $E_\nu$ at the saddle, 
the number $n_u$ of unstable directions 
(excluding the one along the reaction coordinate),
the Lyapunov exponent $\lambda_r$ along the reaction coordinate,
the critical exponent $\mu$ near threshold, and some comment
on the states.}
\begin{center}
\begin{tabular}{rrrrrl}
\hline
$\nu$ & $E_\nu$ & $n_u$ & $\lambda_r$ & $\mu$ & comment\\
\hline
 1 & $-14.8004$ &  4 &   0.9817 &   5.7342 &  all on a ring \\ 
 2 & $-14.8001$ &  4 &   0.9750 &   5.6633 &  ring plus center \\ 
 3 & $-14.7763$ &  5 &   0.9766 &   6.0823 &  --- \\ 
 4 & $-14.3922$ &  6 &   0.9684 &   7.8093 &  --- \\ 
 5 & $-13.5302$ &  7 &   0.9568 &  10.7953 &  all on a line \\ 
\hline
\end{tabular}
\end{center}
\end{table}

\subsection{Six electrons}
\begin{figure}
\begin{center}
\includegraphics*[width=8.6cm]{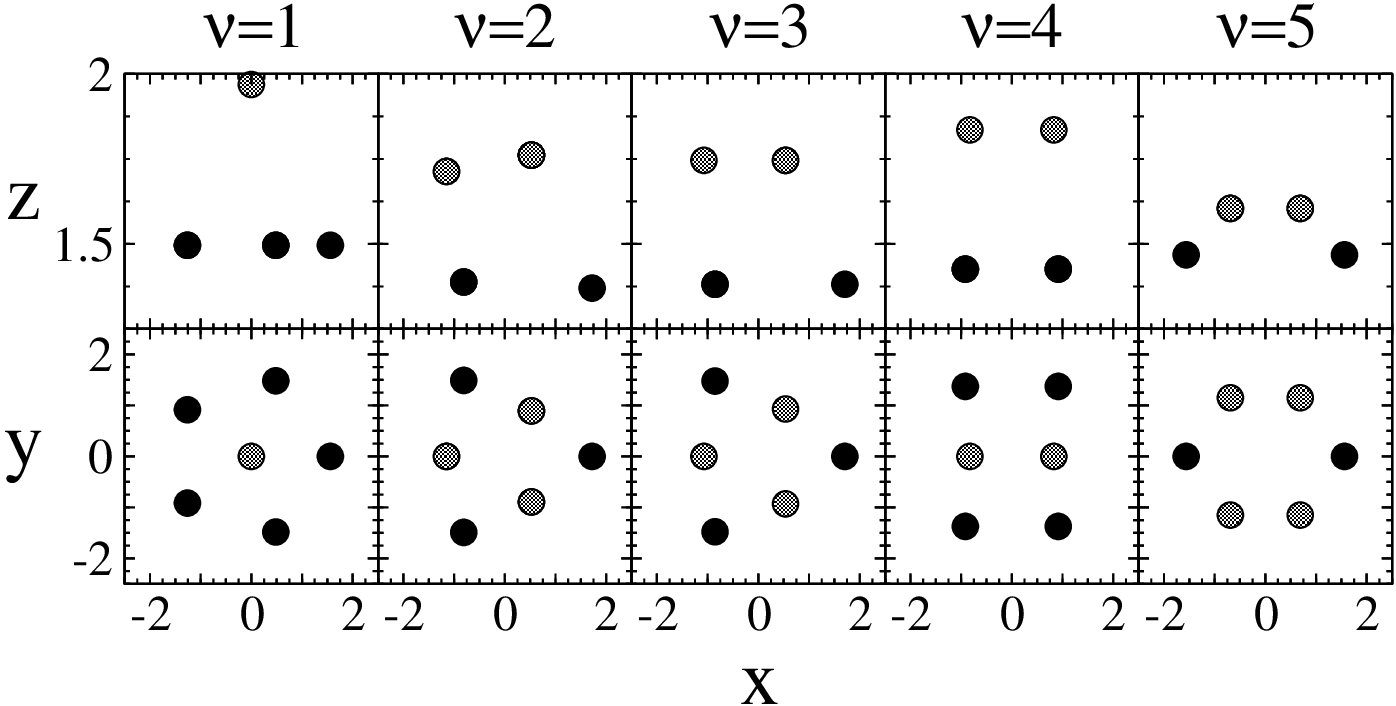}
\end{center}
\caption[]{\label{six_select_fig}
The lowest lying saddles for six electrons, projected
onto the $x$-$z$-plane with the field axis (top row)
and onto the $x$-$y$-plane perpendicular to the field. 
Some electrons are shaded to help identification 
in the two sets of frames.
The states have $C_{5v}$ ($\nu=1$), $C_{v}$ ($\nu=2$), $C_{3v}$ ($\nu=3$) 
and $C_{2v}$ ($\nu=4,5$) symmetries.
}
\end{figure}

For six electrons we find 11 saddles, with properties listed
in table~\ref{all_six}. The first five configurations are
shown in Fig.~\ref{six_select_fig}. The state with 
all particles on a ring is no longer dominant, it appears in the 
middle of table~\ref{all_six}. The saddle with an electron in the center and
the others on a ring dominates the non-sequential ionization process. It has 
both the lowest energy and the smallest critical exponent. 
The states $\nu=2$ and $\nu=3$ look similar, but differ in their
symmetries: state $\nu=2$ has 
a $C_v$ symmetry, state $\nu=3$ has a $C_{3v}$ symmetry.

\begin{table}
\caption[]{\label{all_six}
Saddles for six electrons, ordered by increasing saddle energy.
The columns give the number $\nu$ of the state, the energy $E_\nu$ 
at the saddle, the number $n_u$ of unstable directions 
(excluding the one along the reaction coordinate),
the Lyapunov exponent $\lambda_r$ along the reaction coordinate,
the critical exponent $\mu$ near threshold, and some comment
on the states.}
\begin{center}
\begin{tabular}{rrrrrl}
\hline
$\nu$ & $E_\nu$ & $n_u$ & $\lambda_r$ & $\mu$ & comment\\
\hline
 1 & $-18.8975$ &  5 &   0.9345 &   7.2035 & ring plus center  \\ 
 2 & $-18.7634$ &  6 &   0.9350 &   7.7165 & ---  \\ 
 3 & $-18.7633$ &  7 &   0.9350 &   7.7830 & ---  \\ 
 4 & $-18.7490$ &  6 &   0.9342 &   7.9740 & ---  \\ 
 5 & $-18.7055$ &  7 &   0.9402 &   8.5282 & ---  \\ 
 6 & $-18.7044$ &  8 &   0.9409 &   8.6800 & all on a ring  \\ 
 7 & $-18.6476$ &  7 &   0.9293 &   8.6653 & ---  \\ 
 8 & $-18.5511$ &  7 &   0.9297 &   9.0837 & ---  \\ 
 9 & $-18.2526$ &  8 &   0.9241 &  10.2688 & ---  \\ 
10 & $-17.9228$ &  8 &   0.9205 &  11.6055 & ---  \\ 
11 & $-16.7566$ &  9 &   0.9082 &  15.4282 & all on a line  \\ 
\hline
\end{tabular}
\end{center}
\end{table}

\subsection{Seven electrons}
\begin{figure}
\begin{center}
\includegraphics*[width=8.6cm]{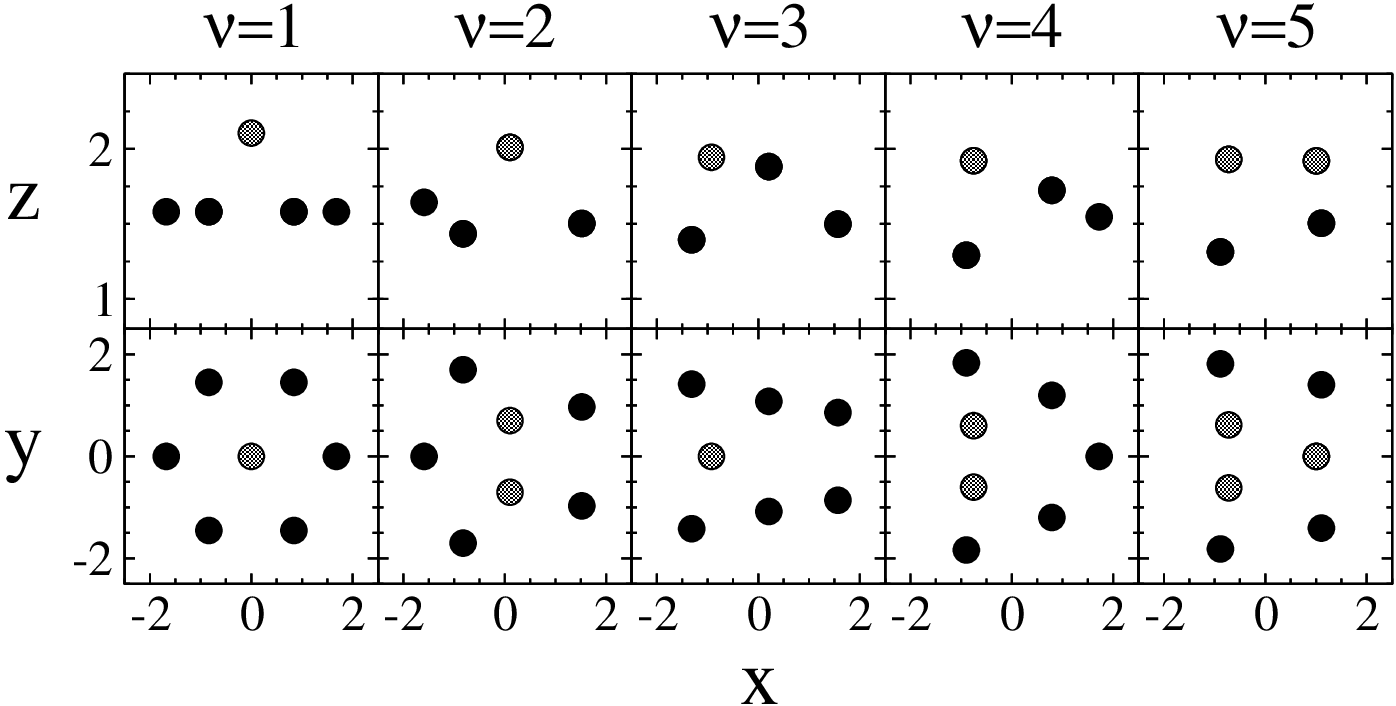}
\end{center}
\caption[]{\label{seven_select_fig}
The lowest lying saddles for seven electrons, projected onto the $x$-$z$-plane 
with the field axis (top row) and onto the 
$x$-$y$-plane perpendicular to the field. 
Some electrons are shaded to help identification 
in the two sets of frames.
The states possess $C_{6v}$ ($\nu=1$) and $C_{v}$ ($\nu=2,3,4,5$) symmetries.
}
\end{figure}

For seven electrons we have 14 states, with the properties listed
in table~\ref{all_seven}. In Fig.~\ref{seven_select_fig} the first five states
are shown. The configuration with one electron in the center and the others 
on a ring dominates the simultaneous electron escape, just as for the six 
electron problem.

\begin{table}
\caption[]{\label{all_seven}
Saddles for seven electrons, ordered by increasing saddle energy.
The columns give the number $\nu$ of the state, the energy $E_\nu$ 
at the saddle, the number $n_u$ of unstable directions 
(excluding the one along the reaction coordinate),
the Lyapunov exponent $\lambda_r$ along the reaction coordinate,
the critical exponent $\mu$ near threshold, and some comment
on the states.}
\begin{center}
\begin{tabular}{rrrrrl}
\hline
$\nu$ & $E_\nu$ & $n_u$ & $\lambda_r$ & $\mu$ & comment\\
\hline
 1 & $-23.1867$ &  6 &   0.9015 &   8.8984 & ring plus center  \\ 
 2 & $-23.0773$ &  7 &   0.8980 &   9.3618 & ---  \\ 
 3 & $-22.9904$ &  8 &   0.8991 &  10.1259 & ---  \\ 
 4 & $-22.8409$ &  8 &   0.8980 &  10.7743 & ---  \\ 
 5 & $-22.8360$ &  8 &   0.8967 &  10.7509 & ---  \\ 
 6 & $-22.8126$ &  9 &   0.9036 &  11.3732 & ---  \\ 
 7 & $-22.7859$ & 10 &   0.9077 &  12.0849 & all on a ring  \\ 
 8 & $-22.7451$ &  9 &   0.8932 &  11.2920 & ---  \\ 
 9 & $-22.5550$ &  9 &   0.8904 &  12.3631 & ---  \\ 
10 & $-22.4109$ &  9 &   0.8905 &  12.8292 & ---  \\ 
11 & $-22.0806$ & 10 &   0.8861 &  13.8150 & ---  \\ 
12 & $-22.0804$ & 11 &   0.8861 &  13.8850 & ---  \\ 
13 & $-21.5215$ & 10 &   0.8805 &  16.1329 & ---  \\ 
14 & $-20.0549$ & 11 &   0.8680 &  20.7764 & all on a line  \\ 
\hline
\end{tabular}
\end{center}
\end{table}

\subsection{Eight electrons}
\begin{figure}
\begin{center}
\includegraphics*[width=8.6cm]{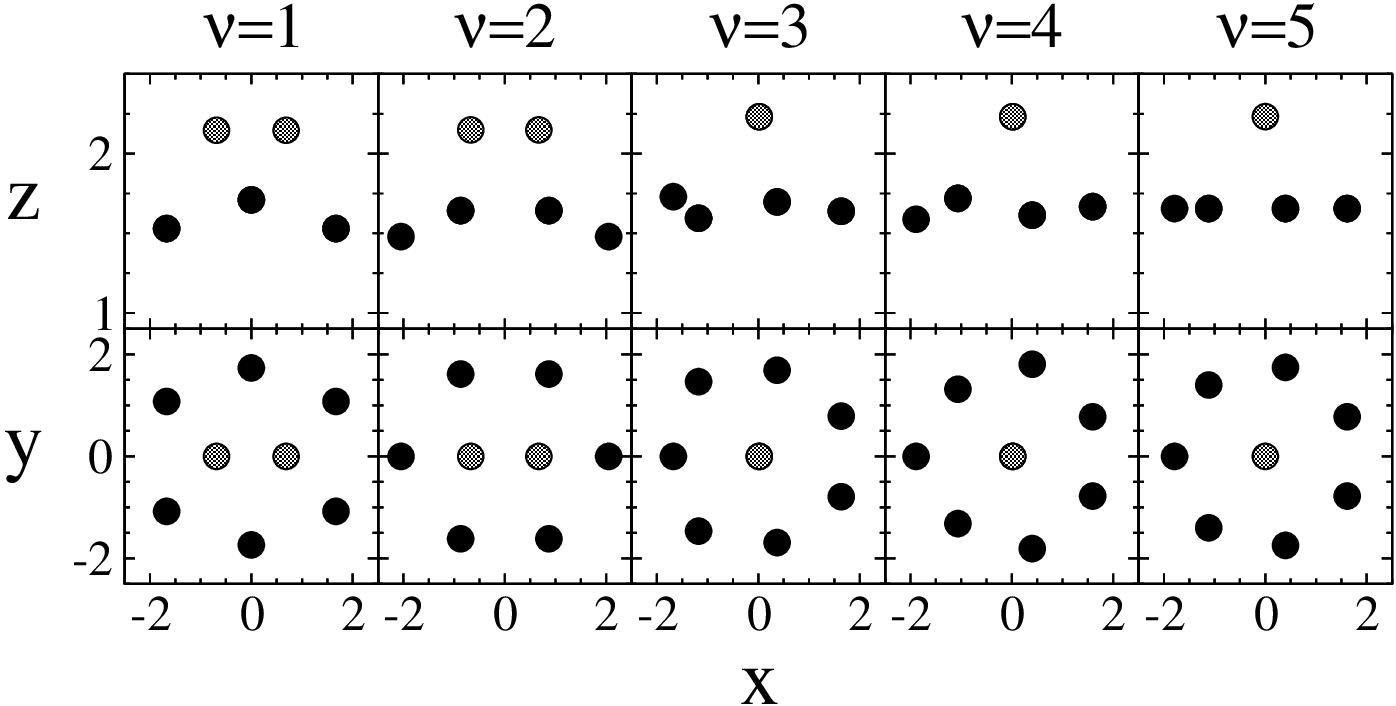}
\end{center}
\caption[]{\label{eight_select_fig}
The lowest lying saddles for eight electrons, projected
onto the $x$-$z$-plane with the field axis (top row)
and onto the $x$-$y$-plane perpendicular to the field. 
Some electrons are shaded to help identification 
in the two sets of frames.
The states possess $C_{2v}$ ($\nu=1,2$), $C_{v}$ ($\nu=3,4$) and $C_{7v}$
($\nu=5$) symmetries.
}
\end{figure}

For eight electrons we have 26 configurations, listed in 
table~\ref{all_eight}. The dominant configuration has two electrons 
in the center and the other six around it, see Fig.~\ref{eight_select_fig}.
The $\nu=2$ saddle also has two electrons in the center, but different
arrangements of the outer electrons.

The state with one electron in the center and the others on a symmetric ring is
the fifth configuration, the one with all electrons
on a ring the 17th configuration.
The saddles $\nu=3$ and $\nu=4$ are
similar to the $\nu=5$ one but they only have a $C_v$ symmetry, not the $C_{7v}$
one as for state $\nu=5$. All of these configurations possess
similar characteristics, see table~\ref{all_eight}.

The state with all particles on a line has the highest energy and the largest
critical exponent. This is the case not only for the eight electron problem but 
for all $N$ values analyzed here.

\begin{table}
\caption[]{\label{all_eight}
Saddles for eight electrons, ordered
by increasing saddle energy.
The columns give an index $\nu$ of the state,
the energy $E_\nu$ at the saddle, 
the number $n_u$ of unstable directions 
(excluding the one along the reaction coordinate),
the Lyapunov exponent $\lambda_r$ along the reaction coordinate,
the critical exponent $\mu$ near threshold, and some comment
on the states.}
\begin{center}
\begin{tabular}{rrrrrl}
\hline
$\nu$ & $E_\nu$ & $n_u$ & $\lambda_r$ &$\mu$ & comment\\
\hline
 1 & $-27.6592$ &  7 &   0.8699 &  10.6553 & two in the center \\ 
 2 & $-27.6523$ &  8 &   0.8698 &  10.9875 & ---  \\ 
 3 & $-27.6373$ &  7 &   0.8737 &  10.7629 & ---  \\ 
 4 & $-27.6373$ &  8 &   0.8737 &  10.7908 & ---  \\ 
 5 & $-27.6372$ &  9 &   0.8739 &  10.9698 & ring plus center  \\ 
 6 & $-27.6363$ &  8 &   0.8724 &  10.9400 & ---  \\ 
 7 & $-27.5495$ &  9 &   0.8679 &  11.5053 & ---  \\ 
 8 & $-27.5125$ &  8 &   0.8683 &  11.4266 & ---  \\ 
 9 & $-27.5122$ &  9 &   0.8684 &  11.5939 & ---  \\ 
10 & $-27.4471$ &  9 &   0.8667 &  12.1438 & ---  \\ 
11 & $-27.2884$ & 10 &   0.8677 &  12.9280 & ---  \\ 
12 & $-27.2785$ & 10 &   0.8692 &  13.1026 & ---  \\ 
13 & $-27.2537$ & 10 &   0.8650 &  13.4763 & ---  \\ 
14 & $-27.2015$ & 10 &   0.8665 &  13.4391 & ---  \\ 
15 & $-27.1801$ & 11 &   0.8629 &  13.4369 & ---  \\ 
16 & $-27.0793$ & 11 &   0.8741 &  14.7803 & ---  \\ 
17 & $-27.0208$ & 12 &   0.8799 &  15.9050 & all on a ring  \\ 
18 & $-26.9869$ & 10 &   0.8639 &  14.3969 & --- \\ 
19 & $-26.9786$ & 10 &   0.8652 &  14.4260 & ---  \\ 
20 & $-26.9774$ & 11 &   0.8659 &  14.7100 & ---  \\ 
21 & $-26.9033$ & 11 &   0.8608 &  14.9062 & ---  \\ 
22 & $-26.5081$ & 11 &   0.8568 &  16.7984 & ---  \\ 
23 & $-26.3255$ & 11 &   0.8567 &  17.2700 & ---  \\ 
24 & $-26.0390$ & 12 &   0.8534 &  18.2821 & --- \\ 
25 & $-25.1700$ & 12 &   0.8463 &  21.3654 & ---  \\ 
26 & $-23.4091$ & 13 &   0.8339 &  26.8366 & all on a line  \\ 
\hline
\end{tabular}
\end{center}
\end{table}

\section{Concluding remarks}
\label{sec4}

The analysis presented here confirms a trend anticipated from our
analysis of triple ionization: that with an increasing number
of electrons more and more saddle configurations and thus
channels for non-sequential multiple ionization have to be
considered. But the analysis also shows that, except for the somewhat
marginal case of fivefold ionization, there is a well separated 
lowest lying saddle that dominates non-sequential
multiple ionization.
Table~\ref{summary_tab} collects the symmetries and the
threshold exponents of this lowest lying state.

A few experiments with multiple ionization of atoms have already 
been performed. However, momenta of the ionization products 
have been measured for double and
triple ionizations only. Currently, measurements of 
momenta of more than two electrons are not available, 
not even of a single component, say the one along the field axis.
But already the measurement of parallel momenta for just two
electrons would suffice for a preliminary test of the 
ionization pathways considered here. In particular, the
breaking of symmetry between the outgoing electrons should
be accessible. As long as the electrons are
interchangeable, the distribution of parallel electron momenta 
should be like that for double ionization,
with a clear preference for similar momenta.
For $N=5$ and more escaping electrons additional structures
in the momentum distribution should appear, since 
the lowest lying saddle has non-equivalent electrons,
see Figs.~\ref{all_four_fig} to \ref{eight_select_fig}.
The momentum distributions for the two
$N=3$ saddles are discussed in \cite{eckhardt01pra2},
and similar modifications in shape can be expected for more
electrons.

The analysis presented here assumes simultaneous escape of all electrons
in a multiple ionization event. It is of course also possible that 
first one or a few electrons escape, then the next batch of electrons leaves
and so on, before one arrives at the final multiply charge ion. 
Sequential ionization becomes significant when the field
intensity is high enough to ionize electrons from a highly excited
state of the ion left behind when the first electrons are gone.
Nevertheless, close to the threshold for  
non-sequential multiple ionization the characteristics
of the non-sequential process as described here should be
identifiable.

\begin{table}
\caption[]{\label{summary_tab}Saddles with lowest energy for different
number of electrons.
The columns give the number $N$ of electrons, 
the energy $E_N$ at the saddle, 
the critical exponent $\mu$ near threshold, the total number of states
$\nu_N$ and some comment on the states.}
\begin{center}
\begin{tabular}{rrrrl}
\hline
$N$ & $E_N$ & $\mu$ & $\nu_N$ & comment\\
\hline
 2 & $ -4.5590$ & 1.2918 & 1 & all on a ring \\ 
 3 & $ -7.6673$ & 2.6226 & 2 & all on a ring \\ 
 4 & $-11.1059$ & 4.0971 & 4 & all on a ring \\ 
 5 & $-14.8004$ & 5.7342 & 5 & all on a ring \\ 
 6 & $-18.8975$ & 7.2035 & 11 & ring plus center \\ 
 7 & $-23.1867$ & 8.8984 & 14 & ring plus center  \\ 
 8 & $-27.6592$ & 10.6553 & 26 & two in the center \\ 
\hline
\end{tabular}
\end{center}
\end{table}

\section{acknowledgment}
We thank J.~M. Rost for a remark that triggered this investigation.
This work was partially supported by the Alexander von Humboldt 
Foundation, the Deutsche Forschungsgemeinschaft and the KBN through grant 
5 P03B 088 21.


\begin{thebibliography}{10}

\bibitem{luk83}
T.~S. Luk, H.~Pummer, K.~Boyer, M.~Shakidi, H.~Egger, and C.~K. Rhodes, 
Phys. Rev. Lett. {\bf 51},  110  (1983).

\bibitem{huillier83}
A. L'Huillier, L.~A. Lompre, G. Mainfray, and C. Manus, Phys. Rev. A {\bf 27},
  2503  (1983).

\bibitem{boyer84}
K.~Boyer, H.~Egger, T.~S. Luk, H.~Pummer, and C.~K. Rhodes, 
J. Opt. Soc. Am. B {\bf 1},  4  (1984).

\bibitem{walker94}
B.~Walker, B.~Sheehy, L.~F. DiMauro, P.~Agostini, K.~J. Schafer, and K.~C.
  Kulander, Phys. Rev. Lett. {\bf 73},  1227  (1994).

\bibitem{moshammer00}
R.~Moshammer, B.~Feuerstein, W.~Schmitt, A.~Dorn, C.D. Sch\"oter, H.~Rottke
  J.~Ullrich, C.~Trump, M.~Wittmann, G.~Korn, K.~Hoffmann, and W.~Sandner, 
Phys. Rev. Lett. {\bf 84},  447  (2000).

\bibitem{weber00prl}
Th. Weber, M.~Weckenbrock, A.~Staudte, L.~Spielberger, O.~Jagutzki, V.~Mergel,
  F.~Afaneh, G.~Urbasch, M.~Vollmer, H.~Giessen, and R.~D\"orner, 
Phys. Rev. Lett. {\bf 84},  443  (2000).

\bibitem{weber00jpb}
Th. Weber, M.~Weckenbrock, A.~Staudte, L.~Spielberger, O.~Jagutzki, V.~Mergel,
  F.~Afaneh, G.~Urbasch, M.~Vollmer, H.~Giessen, and R.~D\"orner, 
J. Phys. B: At. Mol. Opt. Phys. {\bf 33},  L128  (2000).

\bibitem{weber00n}
Th. Weber, H.~Giessen, M.~Weckenbrock, G.~Urbasch, A.~Staudte, L.~Spielberger,
  O.~Jagutzki, V.~Mergel, M.~Vollmer, and R.~D\"orner, 
Nature {\bf 405},  658  (2000).

\bibitem{wechenbrock01}
M.~Wechenbrock, M.~Hattass, A.~Czasch, O.~Jagutzki, L.~Schmidt, T.~Weber,
  H.~Roskos, T.~L\"offler, M.~Thomson, and R.~D\"orner, 
J. Phys. B: At. Mol. Opt. Phys. {\bf 34},  L449
  (2001).

\bibitem{feuerstein01}
B.~Feuerstein, R.~Moshammer, D.~Fischer, A.~Dorn, C.~D. Schr\"oter,
  J.~Deipenwisch, J.~R.~Crespo Lopez-Urrutia, C.~H\"ohr, P.~Neumayer,
  J.~Ullrich, H.~Rottke, C.~Trump, M.~Wittmann, G.~Korn, and W.~Sandner, 
Phys. Rev. Lett. {\bf 87},  043003  (2001).

\bibitem{moshammer02}
R.~Moshammer, B.~Feuerstein, J.~Crespo López-Urrutia, J.~Deipenwisch, A.~Dorn,
  D.~Fischer, C.~Höhr, P.~Neumayer, C.~D. Schröter, J.~Ullrich, H.~Rottke,
  C.~Trump, M.~Wittmann, G.~Korn, and W.~Sandner, 
Phys. Rev. A {\bf 65},  035401  (2002).

\bibitem{corkum93}
P.~B. Corkum, Phys. Rev. Lett. {\bf 71},  1994  (1993).

\bibitem{kulander93}
K.~C. Kulander, K.~J. Schafer, and J.~L. Krause,  in {\em Super-Intense
  Laser-Atom Physics}, {\em Proceedings of the NATO Advanced Research Workshop,
  Han-sur-Lesse, Belgium, 1993}, edited by B. Piraux, A. L'Huillier, and K.
  Rz\c{a}\.zewski (Plenum Press, New York, 1993).

\bibitem{eckhardt01}
B. Eckhardt and K. Sacha, Physica Scripta {\bf T90},  185  (2001).

\bibitem{eckhardt01pra1}
K. Sacha and B. Eckhardt, Phys. Rev. A {\bf 63},  043414  (2001).

\bibitem{eckhardt01silap}
K. Sacha and B. Eckhardt,  in {\em Super-Intense Laser-Atom Physics}, {\em
  Proceedings of the NATO Advanced Research Workshop, Han-sur-Lesse, Belgium,
  2000}, edited by B. Piraux and K. Rz\c{a}\.zewski (Kluwer Academic
  Publishers, Dordrecht, 2001), pp.\ 79--83.

\bibitem{eckhardt01pra2}
K. Sacha and B. Eckhardt, Phys. Rev. A {\bf 64},  053401  (2001).

\bibitem{eckhardt01epl}
B. Eckhardt and K. Sacha, Europhys. Lett. {\bf 56},  651  (2001).

\bibitem{numericalrecipes}
W.~H. Press, S.~A. Teukolsky, W.~T. Vetterling, and B.~P. Flannery, {\em
  Numerical Recipes in FORTRAN The art of Scientific Computing} (Cambridge
  University Press, New York, 1995).

\bibitem{wannier53}
G.~H. Wannier, Phys. Rev. {\bf 90},  817  (1953).

\bibitem{rau84}
A.~R.~P. Rau, Phys. Rep. {\bf 110},  369  (1984).

\bibitem{rost98}
J.~M. Rost, Phys. Rep. {\bf 297},  271  (1998).

\bibitem{rost01phe}
J.~M. Rost, Physica E {\bf 9},  467  (2001).

\bibitem{lambropoulos98}
P. Lambropoulos, P. Maragakis, and J. Zhang, Phys. Rep. {\bf 305},  203
  (1998).

\bibitem{silap93}
  {\em Super-Intense Laser-Atom Physics}, {\em Proceedings of the NATO
  Advanced Research Workshop, Han-sur-Lesse, Belgium, 1993}, edited by B.
  Piraux, A. L'Huillier, and K. Rz\c{a}\.zewski (Plenum Press, New York, 1993).

\bibitem{silap00}
  {\em Super-Intense Laser-Atom Physics}, {\em Proceedings of the NATO
  Advanced Research Workshop, Han-sur-Lesse, Belgium, 2000}, edited by B.
  Piraux and K. Rz\c{a}\.zewski (Kluwer Academic Publishers, Dordrecht, 2001).

\bibitem{becker00}
A. Becker and F.~H.~M. Faisal, Phys. Rev. Lett. {\bf 84},  3546  (2000).

\bibitem{kopold00}
R. Kopold, W. Becker, H. Rottke, and W. Sandner, Phys. Rev. Lett. {\bf 85},
  3781  (2000).

\bibitem{lein00}
M. Lein, E.~K.~U. Gross, and V. Engel, Phys. Rev. Lett. {\bf 85},  4707
  (2000).

\bibitem{feuerstein00}
B. Feuerstein, R. Moshammer, and J. Ullrich, J. Phys. B: At. Mol. Opt. Phys.
  {\bf 33},  L823  (2000).

\bibitem{chen01}
J. Chen, J. Liu, L.~B. Fu, and W.~M. Zheng, Phys. Rev. A {\bf 63},  011404(R)
  (2001).

\bibitem{fu01}
L.-B. Fu, J. Liu, J. Chen, and S.-G. Chen, Phys. Rev. A {\bf 63},  043416
  (2001).

\bibitem{fu02}
L.-B. Fu, J. Liu, and S.-G. Chen, Phys. Rev. A {\bf 65},  021406(R)  (2002).

\bibitem{becker02}
A. Becker and F.~H.~M. Faisal, Phys. Rev. Lett. {\bf 89},  193003  (2002).

\end{thebibliography}

\end{document}